\documentclass[12pt]{article}
\usepackage{amssymb}
\usepackage{amsmath}
\usepackage{graphicx}
\usepackage[unicode,bookmarks,bookmarksopen,bookmarksopenlevel=2,colorlinks,linkcolor=blue,citecolor=green]{hyperref}

\input{tcilatex}
\begin{document}

\title{\textbf{Integrable Dispersive Chains and Their Multi-Phase Solutions.}%
}
\author{Michal Marvan$^{1}$, Maxim V. Pavlov$^{2}$ \\
%EndAName
{\small $^{1}$ Department of Mechanics and Mathematics,}\\
[-3pt] {\small Mathematical Institute in Opava,}\\
[-3pt] {\small Silesian university in Opava}\\
[-3pt] {\small Na Rybn\'{\i}\v{c}ku 1, 746 01 Opava, Czech Republic}\\
[-3pt] {\small $^{2}$ Lebedev Physical Institute of Russian Academy of
Sciences,}\\
[-3pt] {\small Leninskij Prospekt 53, 119991 Moscow, Russia }}
\date{\today }
\maketitle

\begin{abstract}
In this paper we construct multi-phase solutions for integrable dispersive
chains associated with the three-dimensional lnearly-degenerate Mikhal\"{e}v
system of first order. These solutions are parameterized by infinitely many
arbitrary parameters. As byproduct we describe multi-phase solutions for
finite component dispersive reductions of these integrable dispersive chains.
\end{abstract}

\tableofcontents

\newpage

\bigskip \bigskip \newpage 

\begin{flushright}
\textit{dedicated to the memory of Alexander M. Samsonov}
\end{flushright}

\section{Introduction}

\label{sec-intro}

Construction of one-phase solutions for dispersive multi-dimensional
nonlinear systems based on the ansatz \textquotedblleft travelling wave
reduction\textquotedblright , i.e. the dependent functions $%
u_{k}(x,t,y,z,...)$ should be replaced by $u_{k}(\theta )$, where the phase $%
\theta =k_{1}x+k_{2}t+k_{3}y+...$ and $k_{m}$ are constants. This ansatz
leads to systems of ordinary differential equations. Usually this ansatz
allows to find global solutions, i.e. periodic solutions, which are very
important in different applications in physics and mathematics. For
instance, the two-dimensional Korteweg--de Vries equation or the
three-dimensional Kadomtsev--Petviashvili equation have such solutions.
Moreover, these two equations are integrable by the Inverse Scattering
Transform, i.e. they have infinitely many multi-phase solutions.

Recently the authors of the paper \cite{FKK} showed that also \textit{%
dispersionless} three-dimensional systems can possess infinitely many global
solutions. For instance, the authors considered the Mikhal\"{e}v system (see 
\cite{Mikh})%
\begin{equation}
a_{1,t}=a_{2,x},\text{ \ }a_{1}a_{2,x}+a_{1,y}=a_{2}a_{1,x}+a_{2,t},
\label{three}
\end{equation}%
and constructed several explicit global solutions for this system. This
system is integrable by the method of hydrodynamic reductions (see \cite%
{maksjmp}). Its Lax pair is (see again \cite{maksjmp})%
\begin{equation}
p_{t}=[(\lambda +a_{1})p]_{x},\text{ \ }p_{y}=[(\lambda ^{2}+a_{1}\lambda
+a_{2})p]_{x},  \label{pi}
\end{equation}%
where $\lambda $ is an arbitrary parameter. The method of hydrodynamic
reductions means (see \cite{FerKar}) that two families of commuting
hydrodynamic type systems ($N$ is an arbitrary natural number)%
\begin{equation}
r_{t}^{i}=(r^{i}+a_{1})r_{x}^{i},\text{ \ }%
r_{y}^{i}=[(r^{i})^{2}+a_{1}r^{i}+a_{2}]r_{x}^{i},\text{ }i=1,2,...,N
\label{slab}
\end{equation}%
have infinitely many conservation laws (\ref{pi}), where now (!) $a_{1}(%
\mathbf{r}),a_{2}(\mathbf{r})$ and $p(\lambda ,\mathbf{r})$ such that (see 
\cite{maksjmp})%
\begin{equation*}
a_{1}=\sum_{m=1}^{N}f_{m}^{\prime }(r^{m}),\text{ \ }a_{2}=%
\sum_{m=1}^{N}(r^{m}f_{m}^{\prime }(r^{m})-f_{m}(r^{m}))+\frac{1}{2}%
a_{1}^{2},\text{ \ }p=\exp \sum_{m=1}^{N}\int \frac{f_{m}^{\prime \prime
}(r^{m})dr^{m}}{r^{m}-\lambda },
\end{equation*}%
where $f_{k}(r^{k})$ are arbitrary functions. In the particular case ($%
\alpha _{k}$ and $\gamma _{m}$ are arbitrary constants)%
\begin{equation*}
f_{m}(r^{m})=-\frac{1}{2}(r^{m})^{2}+\alpha _{m}r^{m}+\gamma _{m},
\end{equation*}%
hydrodynamic type systems are called linearly-degenerate (i.e. their
characteristic velocities do not depend on corresponding Riemann invariants;
see, for instance \cite{Makslin}). Then (for simplicity here we denote $\eta
_{1}=\Sigma \alpha _{m}$ and $\eta _{2}=\eta _{1}^{2}/2-\Sigma \gamma _{m}$)%
\begin{equation*}
a_{1}=-\sum_{m=1}^{N}r^{m}+\eta _{1},\text{ \ }a_{2}=\frac{1}{2}\left(
\sum_{m=1}^{N}r^{m}\right) ^{2}-\frac{1}{2}\sum_{m=1}^{N}(r^{m})^{2}-\eta
_{1}\sum_{m=1}^{N}r^{m}+\eta _{2},
\end{equation*}%
and%
\begin{equation*}
p=\overset{N}{\underset{m=1}{\prod }}(r^{m}-\lambda )^{-1}.
\end{equation*}%
Just in this particular case, hydrodynamic type systems (\ref{slab}) are
compatible with (see \cite{Fer91} and \cite{Makslin})%
\begin{equation}
r_{x}^{k}=\frac{g_{k}(r^{k})}{\prod_{m\neq k}(r^{k}-r^{m})},\text{ }%
k=1,2,...,N,  \label{arka}
\end{equation}%
where $g_{k}(r^{k})$ are arbitrary functions. Then the extended system (see (%
\ref{slab}))%
\begin{equation*}
r_{x}^{k}=\frac{g_{k}(r^{k})}{\prod_{m\neq k}(r^{k}-r^{m})},\text{ \ }%
r_{t}^{k}=\frac{(r^{k}+a_{1})g_{k}(r^{k})}{\prod_{m\neq k}(r^{k}-r^{m})},%
\text{ \ }r_{y}^{k}=\frac{[(r^{k})^{2}+a_{1}r^{k}+a_{2}]g_{k}(r^{k})}{%
\prod_{m\neq k}(r^{k}-r^{m})}
\end{equation*}%
is integrable in quadratures (see \cite{Fer91}). Then integration of the
above equations leads to the implicit solution%
\begin{equation}
x+\eta _{1}t+\eta _{2}y=\sum_{n=1}^{N}\int^{r^{n}}\frac{\lambda
^{N-1}d\lambda }{g_{n}(\lambda )},\text{ \ }t+\eta
_{1}y=\sum_{n=1}^{N}\int^{r^{n}}\frac{\lambda ^{N-2}d\lambda }{g_{n}(\lambda
)},  \label{gen}
\end{equation}%
\begin{equation*}
y=\sum_{n=1}^{N}\int^{r^{n}}\frac{\lambda ^{N-3}d\lambda }{g_{n}(\lambda )},%
\text{ \ }\pi _{k}=\sum_{n=1}^{N}\int^{r^{n}}\frac{\lambda ^{N-k-1}d\lambda 
}{g_{n}(\lambda )},\text{ }k=3,...,N-1,
\end{equation*}%
where $\pi _{k}$ are arbitrary constants.

Our target in this paper is to select such functions $g_{k}(\lambda )$, that
these solutions simultaneously are multi-phase solutions of the so called $M$%
th dispersive integrable chains (see \cite{energy}) and their infinitely
many dispersive integrable reductions.

So, in this paper we consider integrable dispersive chains (see details in 
\cite{energy})%
\begin{equation}
u_{k,t}=u_{k+1,x}-\frac{1}{2}u_{1}u_{k,x}-u_{k}u_{1,x}+\frac{1}{4}\delta
_{k}^{M}u_{1,xxx},\text{ \ }k=1,2,...  \label{mchains}
\end{equation}%
for all natural numbers $M$. Here $u_{k}$ are unknown functions (or field
variables) and $\delta _{m}^{k}$ is the Kronecker delta. The simplest
example ($M=1$) is the \textit{Korteweg--de Vries chain}%
\begin{equation*}
u_{1,t}=u_{2,x}-\frac{3}{2}u_{1}u_{1,x}+\frac{1}{4}u_{1,xxx},
\end{equation*}%
\begin{equation*}
u_{k,t}=u_{k+1,x}-\frac{1}{2}u_{1}u_{k,x}-u_{k}u_{1,x},\text{ \ }k=2,3,...
\end{equation*}%
The simplest constraint $u_{2}=0$ yields the remarkable Korteweg--de Vries
equation%
\begin{equation}
u_{1,t}=-\frac{3}{2}u_{1}u_{1,x}+\frac{1}{4}u_{1,xxx}.  \label{kdv}
\end{equation}%
The next constraint $u_{3}=0$ implies another well-known integrable system:
the Ito system%
\begin{equation}
u_{1,t}=u_{2,x}-\frac{3}{2}u_{1}u_{1,x}+\frac{1}{4}u_{1,xxx},\text{ \ }%
u_{2,t}=-\frac{1}{2}u_{1}u_{2,x}-u_{2}u_{1,x}.  \label{ito}
\end{equation}%
If $M=2$, the constraint $u_{3}=0$ leads to the Kaup--Boussinesq system%
\begin{equation*}
u_{1,t}=u_{2,x}-\frac{3}{2}u_{1}u_{1,x},\text{ \ }u_{2,t}=-\frac{1}{2}%
u_{1}u_{2,x}-u_{2}u_{1,x}+\frac{1}{4}u_{1,xxx}
\end{equation*}%
connected with the nonlinear Schr\"{o}dinger equation by an appropriate
transformation (see, for instance, \cite{SAlber}).

Multi-phase solutions for these integrable systems are associated with
hyperelliptic Riemann surfaces. $N$ phase solutions of the Korteweg--de
Vries equation are parameterised by $2N+1$ arbitrary constants\footnote{%
Here we ignore phase shifts.}, $N$ phase solutions of the Ito system and of
the nonlinear Schr\"{o}dinger equation are parameterised by $2N+2$ arbitrary
constants. In comparison with finite-component reductions, $N$ phase
solutions of the dispersive integrable chains depend on \textit{infinitely
many} arbitrary parameters. In this paper we utilise a simple construction
of multi-phase solutions based on several first original papers where the
finite-gap integration was established, i.e. \cite{Novikov}, \cite{Dubrovin}%
, \cite{ItsMatveev}. For our convenience we describe these multi-phase
solutions simultaneously for first $N-1$ commuting flows belonging to
corresponding integrable dispersive hierarchies.

In 1919 French mathematician Jules Drach (see \cite{Drach}) formulated the
following problem\footnote{%
The translation of these both papers to English language can be found in 
\cite{Chud} from page 445, to Russian language can be found in \cite{YuB}
from page 20.}: let us consider the linear equation%
\begin{equation*}
\psi _{xx}=\left( \lambda +u\right) \psi ,
\end{equation*}%
where the function $\psi $ depends on an independent variable $x$ and on an
arbitrary parameter $\lambda $, while the function $u$ depends just on an
independent variable $x$. Under the substitution $\psi =\exp \left( \int
qdx\right) $, one can obtain well-known Riccati equation%
\begin{equation*}
q_{x}+q^{2}=\lambda +u.
\end{equation*}%
The Goal: to select all functions $u(x)$ such that this Riccati equation
(with an arbitrary parameter $\lambda $!) is \textit{integrable in
quadratures}.

Jules Drach derived the system of ordinary differential equations (cf. (\ref%
{arka}))%
\begin{equation}
r_{x}^{k}=2\frac{\sqrt{S(r^{k})}}{\prod_{m\neq k}(r^{k}-r^{m})},\text{ }%
k=1,2,...,N,  \label{drach}
\end{equation}%
where $u=2\Sigma r^{n}-\Sigma \beta _{m}$ and ($\beta _{k}$ are arbitrary
parameters)%
\begin{equation*}
S(\lambda )=\overset{2N+1}{\underset{m=1}{\prod }}(\lambda -\beta _{m}).
\end{equation*}%
Then he presented a general solution of this system (\ref{drach}) in
quadratures\footnote{%
The solution found by Jules Drach is exactly coincides with (\ref{gen}), if
to fix $t$ and $y$ to constants.}. These results of Jules Drach were
forgotten for many years. The Korteweg--de Vries equation is determined by
the Lax pair%
\begin{equation*}
\psi _{xx}=\left( \lambda +u\right) \psi ,\text{ \ }\psi _{t}=\left( \lambda
-\frac{1}{2}u\right) \psi _{x}+\frac{1}{4}u_{x}\psi .
\end{equation*}%
Its solution in a similar form was found independently by B.A. Dubrovin in
1975, but also evolution with respect to the time variable $t$ was presented
in \cite{Dubrovin}.

In this paper we select all functions $u_{k}(x)$ such that the Riccati
equation ($M=0,1,2,...$)%
\begin{equation*}
q_{x}+q^{2}=\lambda ^{M}\left( 1+\frac{u_{1}}{\lambda }+\frac{u_{2}}{\lambda
^{2}}+\frac{u_{3}}{\lambda ^{3}}+...\right)
\end{equation*}%
is \textit{integrable in quadratures}. We find dependencies $u_{k}(\mathbf{r}%
)$, where the functions $r^{k}(x)$ satisfy system (\ref{drach}) again as in
the case of the Korteweg--de Vries equation. However, in this case the
function $S(\lambda )$ depends on infinitely many arbitrary parameters%
\footnote{%
Integrability in quadratures of this Riccati equation (except the case $M=0$%
) was considered in \cite{AdlerShabat}, \cite{ShabatJNMP}. However, the
authors presented their construction \textit{without} integration constants,
which play a significant role in description of multi-phase solutions.}.

Three-dimensional integrable linearly-degenerate systems possess
simultaneously hydrodynamic and dispersive reductions (see \cite{lma}, \cite%
{mas}, \cite{mas2}, \cite{maksjmp}, \cite{energy}), while three-dimensional
integrable quasilinear systems possess just infinitely many hydrodynamic
reductions, see details in \cite{gibtsar}, \cite{FerKar}.

For $M$th dispersive integrable chains associated with the Mikhal\"{e}v
system we construct infinitely many multi-phase solutions, which depend on
arbitrary number of arbitrary parameters, but these $N$ phase solutions also
depend on two non-negative integers $M$ and $K$. This is a biggest
difference with two-dimensional integrable dispersive systems: while they
have just one $N$ phase solution for every natural number $N$, $M$th
dispersive integrable chains associated with the Mikhal\"{e}v system possess 
$N$ phase solutions for every pair of non-negative integers $M$ and $K$.
This situation is similar to the famous Kadomtsev--Petviashvili equation,
which also has infinitely many multi-phase solutions (i.e. infinitely many
one-phase solutions, infinitely many two-phase solutions, etc.). The crucial
difference between the Mikhal\"{e}v system and the Kadomtsev--Petviashvili
equation is that the first of them is dispersionless, while the second one
is dispersive\footnote{%
This means that one can widely use the method of hydrodynamic reductions 
\cite{gibtsar}, \cite{FerKar} and the generalised hodograph method \cite%
{Tsar} for the Mikhal\"{e}v system.}. Nevertheless, both of them possess
infinitely many dispersive reductions, like the Korteweg--de Vries equation
and the Kaup--Boussinesq system. Since such two-dimensional integrable
dispersive systems have multi-phase solutions, we can compute multi-phase
solutions for every dispersive reduction of $M$th dispersive integrable
chains associated with the Mikhal\"{e}v system. Thus, they are
simultaneously are multi-phase solutions for the Mikhal\"{e}v system (\ref%
{gen}).

The paper is organised as follows: we discuss integrable dispersive chains (%
\ref{mchains}), and their common properties like conservation laws and
commuting flows in Section \ref{sec:chains}. Then in Section \ref{sec:high}
we construct multi-phase solutions for these integrable dispersive chains (%
\ref{mchains}). In Section \ref{sec:multi} we extract infinitely many
integrable dispersive reductions and discuss their multi-phase solutions. In
Section \ref{sec:except} we consider the exceptional case $M=0$. Finally in
Conclusion \ref{sec:final} we discuss ..........

\section{Integrable Dispersive Chains}

\label{sec:chains}

If the function $\psi $ satisfies the pair of linear equations in partial
derivatives%
\begin{equation}
\psi _{xx}=u\psi ,\text{ \ }\psi _{t}=a\psi _{x}-\frac{1}{2}a_{x}\psi ,
\label{sh}
\end{equation}%
the compatibility condition $(\psi _{xx})_{t}=(\psi _{t})_{xx}$ implies the
relationship%
\begin{equation}
u_{t}=\left( -\frac{1}{2}\partial _{x}^{3}+2u\partial _{x}+u_{x}\right) a.
\label{relay}
\end{equation}%
The choice ($M=1,2,...$, see details in \cite{energy})%
\begin{equation}
u=\lambda ^{M}\left( 1+\frac{u_{1}}{\lambda }+\frac{u_{2}}{\lambda ^{2}}+%
\frac{u_{3}}{\lambda ^{3}}+...\right)  \label{you}
\end{equation}%
leads to infinitely many commuting integrable dispersive chains. Here $%
\lambda $ is the so called spectral parameter.

Indeed, one can introduce infinitely many linear equations%
\begin{equation}
\psi _{t_{k}}=a^{(k)}\psi _{x}-\frac{1}{2}a_{x}^{(k)}\psi ,  \label{tklin}
\end{equation}%
where ($k=1,2,...$)%
\begin{equation}
a^{(k)}=\lambda ^{k}+\lambda ^{k-1}a_{1}+...+\lambda a_{k-1}+a_{k}.
\label{aka}
\end{equation}%
For instance, the compatibility condition $(\psi _{xx})_{t_{1}}=(\psi
_{t_{1}})_{xx}$ yields integrable dispersive chains (\ref{mchains})%
\begin{equation*}
u_{k,t_{1}}=u_{k+1,x}-\frac{1}{2}u_{1}u_{k,x}-u_{k}u_{1,x}+\frac{1}{4}\delta
_{k}^{M}u_{1,xxx},\text{ \ }k=1,2,...,
\end{equation*}%
where%
\begin{equation*}
a_{1}=-\frac{1}{2}u_{1};
\end{equation*}%
the compatibility condition $(\psi _{xx})_{t_{2}}=(\psi _{t_{2}})_{xx}$
implies their first higher commuting flows%
\begin{equation}
u_{k,t_{2}}=u_{k+2,x}-\frac{1}{2}u_{1}u_{k+1,x}-u_{k+1}u_{1,x}+\frac{1}{4}%
\delta _{k}^{M-1}u_{1,xxx}  \label{secondo}
\end{equation}%
\begin{equation*}
+a_{2}u_{k,x}+2u_{k}a_{2,x}-\frac{1}{2}\delta _{k}^{M}a_{2,xxx},
\end{equation*}%
where%
\begin{equation*}
a_{2}=\frac{3}{8}u_{1}^{2}-\frac{1}{2}u_{2}-\frac{1}{8}\delta
_{1}^{M}u_{1,xx}.
\end{equation*}%
Next expression (see \cite{energy}) is more complicated:%
\begin{equation*}
a_{3}=-\frac{1}{2}u^{3}+\frac{3}{4}u^{1}u^{2}-\frac{5}{16}(u^{1})^{3}+\frac{1%
}{32}\delta
_{M}^{1}(10u^{1}u_{xx}^{1}+5(u_{x}^{1})^{2}-u_{xxxx}^{1}-4u_{xx}^{2})-\frac{1%
}{8}\delta _{M}^{2}u_{xx}^{1}.
\end{equation*}%
Nevertheless, every $M$th dispersive chain can be written via these
expressions $a_{k}$ in a very compact form (see \cite{energy} again)%
\begin{equation}
a_{k,t_{1}}=a_{k+1,x}+a_{1}a_{k,x}-a_{k}a_{1,x},\text{ }k=1,2,...  \label{ak}
\end{equation}%
Here we emphasise that a corresponding natural number $M$ and higher order
derivatives are hidden inside of differential polynomials $a_{k}(\mathbf{u},%
\mathbf{u}_{x},\mathbf{u}_{xx},...)$. Higher commuting flows also take
hydrodynamic form in these coordinates $a_{m}$. For instance, (\ref{secondo}%
) becomes%
\begin{equation}
a_{k,t_{2}}=a_{k+2,x}+a_{1}a_{k+1,x}+a_{2}a_{k,x}-a_{k}a_{2,x}-a_{k+1}a_{1,x}.
\label{aks}
\end{equation}

\textbf{Remark}: Eliminating the derivative $a_{3,x}$, Mikhal\"{e}v system (%
\ref{three}) can be obtained from first two equations of (\ref{ak}) and the
first equation of (\ref{aks}).

Integrable dispersive chains (\ref{mchains}) as well as their higher
commuting flows have generating equations of conservation laws (see (\ref%
{aka}))%
\begin{equation}
p_{t_{k}}=(a^{(k)}p)_{x},\text{ \ }k=1,2,...  \label{genek}
\end{equation}%
Here the generating function of conservation law densities $p=1/\varphi $,
where the function $\varphi =\psi \psi ^{+}$, and $\psi ,$ $\psi ^{+}$ are
two linearly independent solutions of the first equation in (\ref{sh}) such
that $\psi \psi _{x}^{+}-\psi ^{+}\psi _{x}=1$. Indeed, an equivalent form
of Lax pair (\ref{sh}) is (we remind the function $u$ is determined by (\ref%
{you}))%
\begin{equation}
\varphi _{xxx}=4u\varphi _{x}+2\varphi u_{x},  \label{fieks}
\end{equation}%
\begin{equation}
\varphi _{t_{k}}=a^{(k)}\varphi _{x}-a_{x}^{(k)}\varphi .  \label{fitime}
\end{equation}%
Finally introducing another function $p=1/\varphi $, (\ref{fitime}) takes
the conservative form (\ref{genek}), while the generating function $%
p(\lambda )$ of conservation law densities satisfies the nonlinear equation%
\begin{equation}
(p^{-1})_{xxx}=4u(p^{-1})_{x}+2p^{-1}u_{x}.  \label{nonlinp}
\end{equation}%
If $\lambda \rightarrow \infty $, the asymptotic behaviour of both solutions
of the linear equation (see the first equation in (\ref{sh}) and (\ref{you}))%
\begin{equation*}
\psi _{xx}=\lambda ^{M}\left( 1+\frac{u_{1}}{\lambda }+\frac{u_{2}}{\lambda
^{2}}+\frac{u_{3}}{\lambda ^{3}}+...\right) \psi
\end{equation*}%
is: $\psi \rightarrow \exp (\lambda ^{M/2}x)$, $\psi ^{+}\rightarrow \exp
(-\lambda ^{M/2}x)$. Thus, the asymptotic behaviour of $\varphi $ starts
from the unity\footnote{%
We remind that the function $\varphi $ satisfies to linear equations (\ref%
{fieks}) and (\ref{fitime}). Thus, the function $\varphi $ is determined up
to an arbitrary constant factor. This means that we fix this factor to unity
for our further convenience here.}, i.e.%
\begin{equation}
\varphi =1+\frac{a_{1}}{\lambda }+\frac{a_{2}}{\lambda ^{2}}+\frac{a_{3}}{%
\lambda ^{3}}+\frac{a_{4}}{\lambda ^{4}}+...  \label{fi}
\end{equation}%
Corresponding asymptotic behaviour of generating function of conservation
law densities also starts from the unity, i.e.%
\begin{equation}
p=1+\frac{\sigma _{1}}{\lambda }+\frac{\sigma _{2}}{\lambda ^{2}}+\frac{%
\sigma _{3}}{\lambda ^{3}}+\frac{\sigma _{4}}{\lambda ^{4}}+...,
\label{sigma}
\end{equation}%
where $p\varphi =1$, i.e.%
\begin{equation}
a_{1}+\sigma _{1}=0,\text{ \ }a_{2}+a_{1}\sigma _{1}+\sigma _{2}=0,\text{ \ }%
a_{3}+a_{2}\sigma _{1}+a_{1}\sigma _{2}+\sigma _{3}=0,...  \label{svyaz}
\end{equation}%
Substituting the expansion (\ref{sigma}) into generating equations of
conservation laws (\ref{genek}), one can express infinitely many local
conservation laws ($a_{0}=1$)%
\begin{equation*}
\bigl(\sigma _{k}\bigr)_{t_{m}}=\Bigl(\sum_{n=0}^{m}a_{n}\sigma _{k+m-n}%
\Bigr)_{x},\text{ \ }k,m=1,2,...
\end{equation*}%
Taking into account relationships (\ref{svyaz}), fluxes of these
conservation laws can be expressed via $\sigma _{m}$ only. For instance%
\begin{equation*}
\bigl(\sigma _{k}\bigr)_{t_{1}}=\Bigl(\sigma _{k+1}-\sigma _{1}\sigma _{k}%
\Bigr)_{x},\quad \bigl(\sigma _{k}\bigr)_{t_{2}}=\Bigl(\sigma _{k+2}-\sigma
_{1}\sigma _{k+1}+\bigl(\sigma _{1}^{2}-\sigma _{2}\bigr)\sigma _{k}\Bigr)%
_{x},
\end{equation*}%
\begin{equation*}
\bigl(\sigma _{k}\bigr)_{t_{3}}=\Bigl(\sigma _{k+3}-\sigma _{1}\sigma _{k+2}+%
\bigl(\sigma _{1}^{2}-\sigma _{2}\bigr)\sigma _{k+1}-\bigl(\sigma
_{1}^{3}-2\sigma _{1}\sigma _{2}+\sigma _{3}\bigr)\sigma _{k}\Bigr)_{x}.
\end{equation*}%
Substituting expansion (\ref{sigma}) into (\ref{nonlinp}) allows to find all
conservation law densities $\sigma _{k}$ as differential polynomials with
respect to the field variables $u_{k}$. For instance,%
\begin{equation*}
\sigma _{1}=\frac{1}{2}u_{1},\text{ \ \ }\sigma _{2}=\frac{1}{2}u_{2}-\frac{1%
}{8}u_{1}^{2}+\frac{1}{8}\delta _{1}^{M}u_{1,xx},
\end{equation*}%
\begin{equation*}
\sigma _{3}=\frac{1}{2}u_{3}-\frac{1}{4}u_{1}u_{2}+\frac{1}{16}u_{1}^{3}+%
\frac{1}{32}\delta _{1}^{M}(u_{1,x})^{2}+\frac{1}{8}\left[ \delta
_{2}^{M}u_{1}+\delta _{1}^{M}\left( u_{2}-\frac{3}{4}u_{1}^{2}+\frac{1}{4}%
u_{1,xx}\right) \right] _{xx}.
\end{equation*}

\section{Multi Phase Solutions}

\label{sec:high}

In this section we consider multi-phase solutions for integrable dispersive
chains (\ref{mchains}) and their first $N-2$ commuting flows (see (\ref%
{tklin}) and (\ref{aka})).

As usual, construction of $N$-phase solutions for integrable systems
associated with linear system (\ref{sh}) (see also (\ref{you}), (\ref{aka}))
is based on the crucial observation: the $\psi $ function (see (\ref{sh}))
as well as the field variables $u_{k}$ no longer depend on the time variable 
$t_{N}$ of $N$th commuting flow (\ref{tklin}). This means that we are
looking for an ansatz for the $\psi $ function in the form\footnote{%
Indeed, one can look for the ansatz $\psi
(t_{0},t_{1},...,t_{N-1},t_{N},t_{N+1},...)=\chi (t_{N})\tilde{\psi}%
(t_{0},t_{1},...,t_{N-1})$, where $\chi (t_{N})$ is not yet known function.
Then $N$th equation from (\ref{tklin}) yields, $\frac{\chi ^{\prime }(t_{N})%
}{\chi (t_{N})}\tilde{\psi}=a^{(N)}\tilde{\psi}_{x}-\frac{1}{2}a_{x}^{(N)}%
\tilde{\psi}$. Since the functions $\tilde{\psi}$ and $a^{(N)}$ do not
depend on $t_{N}$, the ratio $\frac{\chi ^{\prime }(t_{N})}{\chi (t_{N})}$
also does not depend on $t_{N}$. This means, that $\chi (t_{N})=\exp (\mu
t_{N})$, where $\mu $ is an arbitrary constant. Below we show that $\tilde{%
\psi}$ also does not depend on higher time variables $t_{N+1},t_{N+2},...$}%
\begin{equation*}
\psi (t_{0},t_{1},...,t_{N-1},t_{N},t_{N+1},...)=e^{\mu t_{N}}\tilde{\psi}%
(t_{0},t_{1},...,t_{N-1}),
\end{equation*}%
where $\mu $ is an arbitrary constant. In this case $N$th equation from (\ref%
{tklin}) becomes%
\begin{equation}
\mu \tilde{\psi}=\tilde{a}^{(N)}\tilde{\psi}_{x}-\frac{1}{2}\tilde{a}%
_{x}^{(N)}\tilde{\psi},  \label{mju}
\end{equation}%
where (see (\ref{aka}))%
\begin{equation}
\tilde{a}^{(N)}=\lambda ^{N}+\lambda ^{N-1}\tilde{a}_{1}+...+\lambda \tilde{a%
}_{N-1}+\tilde{a}_{N}=a^{(N)}+\kappa _{1}a^{(N-1)}+...+\kappa
_{N-1}a^{(1)}+\kappa _{N},  \label{poly}
\end{equation}%
and $\kappa _{m}$ are arbitrary constants. Indeed, corresponding
relationship (\ref{relay})%
\begin{equation*}
u_{t_{N}}=\left( -\frac{1}{2}\partial _{x}^{3}+2u\partial _{x}+u_{x}\right) 
\tilde{a}^{(N)}
\end{equation*}%
reduces to the \textquotedblleft stationary\textquotedblright\ form (with
respect to the time variable $t_{N}$, because also all field variables $%
u_{k} $ do not depend on this time variable $t_{N}$)%
\begin{equation}
-\frac{1}{2}\tilde{a}_{xxx}^{(N)}+2u\tilde{a}_{x}^{(N)}+\tilde{a}%
^{(N)}u_{x}=0.  \label{station}
\end{equation}%
This linear equation on function $\tilde{a}^{(N)}$ is determined up to any
linear combination of lower expressions $\tilde{a}^{(k)}$. Thus, the
parameters $\kappa _{m}$ play an important role in construction of
corresponding $N$-phase solutions.

So, equation (\ref{station}) exactly coincides with (\ref{fieks}), where $%
\varphi \rightarrow \Phi =\tilde{a}^{(N)}$. Thus, the \textquotedblleft
stationary\textquotedblright\ reduction $\psi _{t_{N}}=0$ leads to the
polynomial ansatz (see (\ref{poly}))%
\begin{equation}
\Phi (\lambda ,\mathbf{r})=\overset{N}{\underset{m=1}{\prod }}(\lambda
-r^{m}(x,t))  \label{poli}
\end{equation}%
for linear system (\ref{fieks}), (\ref{fitime}). Taking into account (\ref%
{mju}) the first equation in (\ref{sh}) implies the first integral of (\ref%
{station}):%
\begin{equation}
\mu ^{2}=u(\tilde{a}^{(N)})^{2}+\frac{1}{4}(\tilde{a}_{x}^{(N)})^{2}-\frac{1%
}{2}\tilde{a}^{(N)}\tilde{a}_{xx}^{(N)}.  \label{mukva}
\end{equation}%
Indeed, substituting (see (\ref{mju}))%
\begin{equation*}
\tilde{\psi}_{x}=\left( \frac{\mu }{\tilde{a}^{(N)}}+\frac{1}{2}\frac{\tilde{%
a}_{x}^{(N)}}{\tilde{a}^{(N)}}\right) \tilde{\psi}
\end{equation*}%
into the first equation in (\ref{sh}), we obtain%
\begin{equation*}
\left( \frac{\mu }{\tilde{a}^{(N)}}+\frac{1}{2}\frac{\tilde{a}_{x}^{(N)}}{%
\tilde{a}^{(N)}}\right) \tilde{\psi}_{x}+\left[ \frac{1}{2}\left( \frac{%
\tilde{a}_{x}^{(N)}}{\tilde{a}^{(N)}}\right) _{x}-\mu \frac{\tilde{a}%
_{x}^{(N)}}{(\tilde{a}^{(N)})^{2}}\right] \tilde{\psi}=u\tilde{\psi}.
\end{equation*}%
Finally, eliminating $\tilde{\psi}_{x}$ from both above equations, one can
obtain first integral (\ref{mukva}).

So, we are looking for polynomial solutions (of a degree $N$ with respect to
the spectral parameter $\lambda $) of the \textit{first integral} (\ref%
{fieks})%
\begin{equation}
2\Phi \Phi _{xx}-\Phi _{x}^{2}=4\lambda ^{M}\left( 1+\frac{u_{1}}{\lambda }+%
\frac{u_{2}}{\lambda ^{2}}+\frac{u_{3}}{\lambda ^{3}}+...\right) \Phi
^{2}-4S(\lambda ),  \label{kvadro}
\end{equation}%
where $\mu ^{2}=S(\lambda )$ is an \textquotedblleft integration
constant\textquotedblright , see (\ref{mukva}).

Substitution (\ref{poli}) into (\ref{kvadro}) leads to (see (\ref{you}))%
\begin{equation*}
u(\lambda ,\mathbf{r},\mathbf{r}_{x},\mathbf{r}_{xx})\equiv \lambda
^{M}\left( 1+\frac{u_{1}}{\lambda }+\frac{u_{2}}{\lambda ^{2}}+\frac{u_{3}}{%
\lambda ^{3}}+...\right) =\frac{\Phi _{xx}}{2\Phi }-\frac{\Phi _{x}^{2}}{%
4\Phi ^{2}}+\frac{S(\lambda )}{\Phi ^{2}}
\end{equation*}%
\begin{equation}
=\frac{1}{2}\left( \ln \prod_{n=1}^{N}(\lambda -r^{n})\right) _{xx}+\frac{1}{%
4}\left[ \left( \ln \prod_{n=1}^{N}(\lambda -r^{n})\right) _{x}\right] ^{2}+%
\frac{S(\lambda )}{\prod_{n=1}^{N}(\lambda -r^{n})^{2}},  \label{inter}
\end{equation}%
where ($s_{k}$ are integration constants)%
\begin{equation}
S(\lambda )=\lambda ^{2N+M}\left( 1+\frac{s_{1}}{\lambda }+\frac{s_{2}}{%
\lambda ^{2}}+\frac{s_{3}}{\lambda ^{3}}+...\right) .  \label{sryad}
\end{equation}%
So, we have (\ref{kvadro}) with polynomial ansatz (\ref{poli}) and similar
equation (equivalent to (\ref{fieks}))%
\begin{equation}
2\varphi \varphi _{xx}-\varphi _{x}^{2}=4\lambda ^{M}\left( 1+\frac{u_{1}}{%
\lambda }+\frac{u_{2}}{\lambda ^{2}}+\frac{u_{3}}{\lambda ^{3}}+...\right)
\varphi ^{2}-4\lambda ^{M},  \label{norm}
\end{equation}%
where coefficients $a_{k}$ are determined by (\ref{fi}). Eliminating $u$
from (\ref{kvadro}) and (\ref{norm}), one can obtain%
\begin{equation*}
2\frac{\Phi _{xx}}{\Phi }-\frac{\Phi _{x}^{2}}{\Phi ^{2}}+4\frac{S(\lambda )%
}{\Phi ^{2}}=2\frac{\varphi _{xx}}{\varphi }-\frac{\varphi _{x}^{2}}{\varphi
^{2}}+4\frac{\lambda ^{M}}{\varphi ^{2}}.
\end{equation*}%
This means%
\begin{equation}
\varphi =\lambda ^{-N}\left( 1+\frac{s_{1}}{\lambda }+\frac{s_{2}}{\lambda
^{2}}+\frac{s_{3}}{\lambda ^{3}}+...\right) ^{-1/2}\Phi .  \label{fifi}
\end{equation}%
Thus, all dependencies $a_{k}(\mathbf{r})$ can be found by substitution (\ref%
{fi}) and (\ref{poli}) into the above equation, i.e. ($a_{0}=1$, $\eta
_{0}=1 $)%
\begin{equation}
a_{n}=\sum_{m=0}^{n}\eta _{n-m}\tilde{a}_{m},\text{ }n=1,2,...,N;
\label{belpoli}
\end{equation}%
\begin{equation*}
a_{n}=\sum_{m=0}^{N}\eta _{n-m}\tilde{a}_{m},\text{ }n=N+1,N+2,...,
\end{equation*}%
where parameters $\eta _{m}$ can be found from the expansion%
\begin{equation}
1+\frac{\eta _{1}}{\lambda }+\frac{\eta _{2}}{\lambda ^{2}}+\frac{\eta _{3}}{%
\lambda ^{3}}+...=\left( 1+\frac{s_{1}}{\lambda }+\frac{s_{2}}{\lambda ^{2}}+%
\frac{s_{3}}{\lambda ^{3}}+...\right) ^{-1/2}.  \label{etas}
\end{equation}%
Hence, 
\begin{equation*}
\eta _{p}=\sum_{k_{1},\dots ,k_{m}}\frac{(-1)^{k_{1}+\cdots +k_{m}}}{%
2^{2k_{1}+\cdots +2k_{m}-1}}\frac{(2k_{1}+\cdots +2k_{m}-1)!}{(k_{1}+\cdots
+k_{m}-1)!\,k_{1}!\cdots k_{m}!}s_{i_{1}}^{k_{1}}\cdots s_{i_{m}}^{k_{m}},
\end{equation*}%
where it is assumed that $i_{1},k_{1},\dots ,i_{m},k_{m}$ run over all $2m$%
-tuples such that $i_{1}k_{1}+\cdots +i_{m}k_{m}=p$ and $i_{1},\dots ,i_{m}$
are pairwise different. For instance, 
\begin{equation*}
\begin{aligned} \eta_1 &= -\tfrac{1}{2} s_1, \\ \eta_2 &= -\tfrac{1}{2} s_2
+ \tfrac{3}{8} s_1^2 , \\ \eta_3 &= -\tfrac{1}{2} s_3 + \tfrac{3}{4} s_1 s_2
- \tfrac{5}{16} s_1^3, \\ \eta_4 &= -\tfrac{1}{2} s_4 + \tfrac{3}{4} s_1 s_3
+ \tfrac{3}{8} s_2^2 - \tfrac{15}{16} s_1^2 s_2 + \tfrac{35}{128} s_1^4.
\end{aligned}
\end{equation*}
Also (see (\ref{aka}) and (\ref{poli})) all parameters $\kappa _{m}$ in (\ref%
{poly}) can be found from the inverse formula%
\begin{equation}
1+\frac{\kappa _{1}}{\lambda }+\frac{\kappa _{2}}{\lambda ^{2}}+\frac{\kappa
_{3}}{\lambda ^{3}}+...=\left( 1+\frac{s_{1}}{\lambda }+\frac{s_{2}}{\lambda
^{2}}+\frac{s_{3}}{\lambda ^{3}}+...\right) ^{1/2}.  \label{kapas}
\end{equation}%
Hence, 
\begin{equation*}
\kappa _{p}=\sum_{k_{1},\dots ,k_{m}}\frac{(-1)^{k_{1}+\cdots +k_{m}+1}}{%
2^{2k_{1}+\cdots +2k_{m}-1}}\frac{(2k_{1}+\cdots +2k_{m}-2)!}{(k_{1}+\cdots
+k_{m}-1)!\,k_{1}!\cdots k_{m}!}s_{i_{1}}^{k_{1}}\cdots s_{i_{m}}^{k_{m}}.
\end{equation*}%
Here it is assumed that $i_{1},k_{1},\dots ,i_{m},k_{m}$ run over all $2m$%
-tuples such that $i_{1}k_{1}+\cdots +i_{m}k_{m}=p$ and $i_{1},\dots ,i_{m}$
are pairwise different. For instance,%
\begin{equation*}
\begin{aligned} \kappa_1 &= \tfrac{1}{2} s_1, \\ \kappa_2 &= \tfrac{1}{2}
s_2 - \tfrac{1}{8} s_1^2, \\ \kappa_3 &= \tfrac{1}{2} s_3 - \tfrac{1}{4} s_1
s_2 + \tfrac{1}{16} s_1^3, \\ \kappa_4 &= \tfrac{1}{2} s_4 - \tfrac{1}{4}
s_1 s_3 - \tfrac{1}{8} s_2^2 + \tfrac{3}{16} s_1^2 s_2 - \tfrac{5}{128}
s_1^4 . \end{aligned}
\end{equation*}

Following J. Drach \cite{Drach} and B.A. Dubrovin \cite{Dubrovin}, we
consider the limit $\lambda \rightarrow r^{i}(x,t)$ of (\ref{kvadro}). This
straightforward computation yields the $N$ component system (cf. (\ref{arka}%
) and (\ref{drach}))%
\begin{equation}
r_{x}^{k}=2\frac{\sqrt{S(r^{k})}}{\prod_{m\neq k}(r^{k}-r^{m})},\text{ }%
k=1,2,...,N.  \label{separato}
\end{equation}%
Substituting these expressions back into (\ref{inter}), one can obtain%
\footnote{%
First order derivatives of expressions in parentheses lead to all possible
combinations $r_{x}^{i}r_{x}^{k}$. However, taking into account also second
order derivatives, these products will be cancelled simultaneously.}%
\begin{equation*}
u(\lambda ,\mathbf{r})=\frac{S(\lambda )}{\prod_{n=1}^{N}(\lambda -r^{n})^{2}%
}
\end{equation*}%
\begin{equation*}
+\sum_{n=1}^{N}\frac{1}{\lambda -r^{n}}\left( 2\sum_{m\neq n}\frac{1}{%
r^{n}-r^{m}}-\frac{1}{\lambda -r^{n}}-\frac{S^{\prime }(r^{n})}{S(r^{n})}%
\right) \frac{S(r^{n})}{\prod_{s\neq n}(r^{n}-r^{s})^{2}}.
\end{equation*}%
Taking into account (\ref{sryad}) this equality becomes%
\begin{equation}
u(\lambda ,\mathbf{r})=\frac{S(\lambda )}{\prod_{n=1}^{N}(\lambda -r^{n})^{2}%
}+\sum_{p=0}^{\infty }s_{p}Q_{2N+M-p},  \label{youshort}
\end{equation}%
where ($k=0,\pm 1,\pm 2,...$)%
\begin{equation}
Q_{k}(\lambda ,\mathbf{r})=\sum_{n=1}^{N}\frac{1}{\lambda -r^{n}}\left(
2\sum_{m\neq n}\frac{r^{n}}{r^{n}-r^{m}}-\frac{r^{n}}{\lambda -r^{n}}%
-k\right) \frac{(r^{n})^{k-1}}{\prod_{q\neq n}(r^{n}-r^{q})^{2}}.
\label{short}
\end{equation}

Now we introduce the power sum symmetric polynomials%
\begin{equation*}
c_{k}(\mathbf{r})=\frac{1}{k}\sum_{m=1}^{N}(r^{m})^{k},\text{ \ }c_{k}(%
\mathbf{R})=\frac{1}{k}\sum_{m=1}^{N}(R^{m})^{k},\text{ \ }k=1,2,...,
\end{equation*}%
\textit{where }$R^{n}=1/r^{n}$.

\textbf{Lemma}: \textit{The coefficients} $u_{m}(\mathbf{r})$ \textit{of the
expansion} (\ref{inter})%
\begin{equation*}
u(\lambda ,\mathbf{r})=\lambda ^{M}\left( 1+\frac{u_{1}(\mathbf{r})}{\lambda 
}+\frac{u_{2}(\mathbf{r})}{\lambda ^{2}}+\frac{u_{3}(\mathbf{r})}{\lambda
^{3}}+...\right) ,\text{ }\lambda \rightarrow \infty
\end{equation*}%
\textit{are determined by}%
\begin{equation}
u_{m}(\mathbf{r})=\sum_{k=0}^{m}s_{m-k}\sum_{n=0}^{k}B_{n}B_{k-n},\text{ \ }%
1\leqslant m\leqslant M;  \label{raz}
\end{equation}%
\begin{equation}
u_{m+M}(\mathbf{r})=\sum_{k=1}^{\infty
}s_{2N+M+k+m-1}\sum_{n=1}^{k}B_{-n}B_{n-1-k},\text{ \ }m\geqslant 1,
\label{dva}
\end{equation}%
\textit{where }$B_{k}(\mathbf{r})$ \textit{are Bell polynomials, i.e.}%
\begin{equation}
B(\lambda ,\mathbf{r})=1+\frac{B_{1}(\mathbf{r})}{\lambda }+\frac{B_{2}(%
\mathbf{r})}{\lambda ^{2}}+\frac{B_{3}(\mathbf{r})}{\lambda ^{3}}+...=\exp
\left( \frac{c_{1}(\mathbf{r})}{\lambda }+\frac{c_{2}(\mathbf{r})}{\lambda
^{2}}+\frac{c_{3}(\mathbf{r})}{\lambda ^{3}}+...\right) ;  \label{bel}
\end{equation}%
\begin{equation}
B_{-1}=\underset{m=1}{\overset{N}{\prod }}R^{m}\equiv \underset{m=1}{\overset%
{N}{\prod }}(r^{m})^{-1}.  \label{bminus}
\end{equation}%
\textit{and all other negative }$B_{k}(\mathbf{r})$ \textit{are proportional
to Bell polynomials} \textit{up to the common factor }$B_{-1}$ (\textit{%
expressible at} $\lambda \rightarrow 0$)%
\begin{equation*}
B(\lambda ,\mathbf{r})=B_{-1}+\lambda B_{-2}+\lambda
^{2}B_{-3}+...=B_{-1}\exp (\lambda c_{1}(\mathbf{R})+\lambda ^{2}c_{2}(%
\mathbf{R})+\lambda ^{3}c_{3}(\mathbf{R})+...).
\end{equation*}

\textbf{Examples}:%
\begin{equation}
B_{1}=c_{1}(\mathbf{r}),\text{ \ }B_{2}=c_{2}(\mathbf{r})+\frac{1}{2}%
c_{1}^{2}(\mathbf{r}),\text{ \ }B_{3}=c_{3}(\mathbf{r})+c_{1}(\mathbf{r}%
)c_{2}(\mathbf{r})+\frac{1}{6}c_{1}^{3}(\mathbf{r}),...;  \label{bell}
\end{equation}%
\begin{equation}
B_{-2}=B_{-1}c_{1}(\mathbf{R}),\text{ \ }B_{-3}=B_{-1}\left( c_{2}(\mathbf{R}%
)+\frac{1}{2}c_{1}^{2}(\mathbf{R})\right) ,  \label{bella}
\end{equation}%
\begin{equation*}
B_{-4}=B_{-1}\left( c_{3}(\mathbf{R})+c_{1}(\mathbf{R})c_{2}(\mathbf{R})+%
\frac{1}{6}c_{1}^{3}(\mathbf{R})\right) ,...
\end{equation*}

\textbf{Proof}: At first we introduce the generating function%
\begin{equation*}
Q(\lambda ,\zeta ,\mathbf{r})=\sum_{n=1}^{N}\frac{1}{\lambda -r^{n}}\frac{1}{%
\zeta -r^{n}}\frac{1}{\prod_{q\neq n}(r^{n}-r^{q})^{2}}\left( 2\sum_{m\neq n}%
\frac{1}{r^{n}-r^{m}}-\frac{1}{\lambda -r^{n}}-\frac{1}{\zeta -r^{n}}\right)
,
\end{equation*}%
such that (see (\ref{short}))%
\begin{equation*}
Q(\lambda ,\zeta ,\mathbf{r})=\sum_{k=0}^{\infty }Q_{k}(\lambda ,\mathbf{r}%
)\zeta ^{-k-1},\text{ \ }\zeta \rightarrow \infty ;
\end{equation*}%
\begin{equation*}
Q(\lambda ,\zeta ,\mathbf{r})=-\sum_{k=1}^{\infty }Q_{-k}(\lambda ,\mathbf{r}%
)\zeta ^{k-1},\text{ \ }\zeta \rightarrow 0.
\end{equation*}%
This generating function can be presented in the form%
\begin{equation*}
Q(\lambda ,\zeta ,\mathbf{r})=\frac{G(\lambda ,\mathbf{r})-G(\zeta ,\mathbf{r%
})}{\lambda -\zeta },
\end{equation*}%
where%
\begin{equation*}
G(\lambda ,\mathbf{r})=\sum_{n=1}^{N}\frac{1}{\prod_{q\neq
n}(r^{n}-r^{q})^{2}}\frac{1}{(\lambda -r^{n})^{2}}-2\sum_{n=1}^{N}\frac{1}{%
\prod_{q\neq n}(r^{n}-r^{q})^{2}}\frac{1}{\lambda -r^{n}}\sum_{m\neq n}\frac{%
1}{r^{n}-r^{m}}.
\end{equation*}%
However the above expression is nothing but a partial fraction decomposition
of the product%
\begin{equation*}
G(\lambda ,\mathbf{r})=\prod_{n=1}^{N}(\lambda -r^{n})^{-2}.
\end{equation*}%
Thus, all expressions $Q_{k}(\lambda ,\mathbf{r})$ can be found by expansion
with respect to the parameter $\zeta $ from the generating function%
\begin{equation*}
Q(\lambda ,\zeta ,\mathbf{r})=\frac{1}{\lambda -\zeta }\left( \frac{1}{%
\prod_{n=1}^{N}(\lambda -r^{n})^{2}}-\frac{1}{\prod_{n=1}^{N}(\zeta
-r^{n})^{2}}\right) .
\end{equation*}%
Now we introduce the generating function (see (\ref{poli}))%
\begin{equation*}
B(\lambda ,\mathbf{r})=\frac{1}{\Phi (\lambda ,\mathbf{r})}=\underset{m=1}{%
\overset{N}{\prod }}(\lambda -r^{m})^{-1}.
\end{equation*}%
So, if $\lambda \rightarrow \infty $, then ($B_{0}\equiv 1$)%
\begin{equation}
B(\lambda ,\mathbf{r})=\lambda ^{-N}\left( 1+\frac{B_{1}}{\lambda }+\frac{%
B_{2}}{\lambda ^{2}}+\frac{B_{3}}{\lambda ^{3}}+...\right) ,  \label{bryad}
\end{equation}%
where%
\begin{equation*}
B_{m}=\sum_{n=1}^{N}\frac{(r^{n})^{N+m-1}}{\prod_{s\neq n}(r^{n}-r^{s})}.
\end{equation*}%
These functions $B_{k}(\mathbf{r})$ are Bell polynomials (\ref{bell}).

Then ($\zeta \rightarrow \infty $)%
\begin{equation*}
Q(\lambda ,\zeta ,\mathbf{r})=\frac{B^{2}(\zeta ,\mathbf{r})-B^{2}(\lambda ,%
\mathbf{r})}{\zeta -\lambda }
\end{equation*}%
\begin{equation*}
=\sum_{n=0}^{\infty }\frac{1}{\zeta ^{n+2N+1}}\sum_{m=0}^{n}\left(
\sum_{k=0}^{m}B_{k}B_{m-k}\right) \lambda ^{n-m}-B^{2}(\lambda ,\mathbf{r}%
)\sum_{n=0}^{\infty }\frac{\lambda ^{n}}{\zeta ^{n+1}}.
\end{equation*}%
If $\lambda \rightarrow 0$, then%
\begin{equation*}
B(\lambda ,\mathbf{r})=B_{-1}+\lambda B_{-2}+\lambda ^{2}B_{-3}+...,
\end{equation*}%
where%
\begin{equation*}
B_{-m}=-\sum_{n=1}^{N}\frac{(r^{n})^{-m}}{\prod_{s\neq n}(r^{n}-r^{s})}.
\end{equation*}%
The function $B_{-1}$ is determined by (\ref{bminus}), and the other
functions $B_{-k}(\mathbf{r})$ are determined by (\ref{bella}).

Then ($\zeta \rightarrow 0$)%
\begin{equation*}
Q(\lambda ,\zeta ,\mathbf{r})=\frac{B^{2}(\lambda ,\mathbf{r})-B^{2}(\zeta ,%
\mathbf{r})}{\lambda -\zeta }
\end{equation*}%
\begin{equation*}
=\sum_{m=0}^{\infty }\zeta ^{m}\left[ B^{2}(\lambda ,\mathbf{r})\lambda
^{-m-1}-\sum_{k=0}^{m}\left( \sum_{n=0}^{k}B_{-n-1}B_{n-1-k}\right) \lambda
^{k-m-1}\right] .
\end{equation*}%
The equality (\ref{youshort}) we rewrite in the form%
\begin{equation}
u(\lambda ,\mathbf{r})=\frac{S(\lambda )}{\prod_{n=1}^{N}(\lambda -r^{n})^{2}%
}+\sum_{k=0}^{2N+M}s_{2N+M-k}Q_{k}+\sum_{k=1}^{\infty }s_{2N+M+k}Q_{-k},
\label{youmid}
\end{equation}%
where%
\begin{equation*}
Q_{k}(\lambda ,\mathbf{r})=-B^{2}(\lambda ,\mathbf{r})\lambda ^{k},\text{ \ }%
k=0,1,...,2N-1,
\end{equation*}%
\begin{equation*}
Q_{k}(\lambda ,\mathbf{r})=\sum_{m=0}^{k-2N}\left(
\sum_{n=0}^{m}B_{n}B_{m-n}\right) \lambda ^{k-2N-m}-B^{2}(\lambda ,\mathbf{r}%
)\lambda ^{k},\text{ \ }k=2N,2N+1,...,
\end{equation*}%
\begin{equation*}
Q_{-k}(\lambda ,\mathbf{r})=\sum_{m=0}^{k-1}\left(
\sum_{n=0}^{m}B_{-n-1}B_{n-1-m}\right) \lambda ^{m-k}-B^{2}(\lambda ,\mathbf{%
r})\lambda ^{-k},\text{ }k=1,2,...
\end{equation*}%
Thus (\ref{youmid}) becomes%
\begin{equation*}
u(\lambda ,\mathbf{r})=\sum_{k=0}^{M}s_{M-k}\sum_{m=0}^{k}\left(
\sum_{n=0}^{m}B_{n}B_{m-n}\right) \lambda ^{k-m}
\end{equation*}%
\begin{equation*}
+\sum_{k=1}^{\infty }s_{2N+M+k}\sum_{m=1}^{k}\left(
\sum_{n=1}^{m}B_{-n}B_{n-m-1}\right) \lambda ^{m-k-1}.
\end{equation*}%
Taking into account (\ref{you}) we find (\ref{raz}) and (\ref{dva}).

Again following B.A. Dubrovin \cite{Dubrovin}, we consider the limit $%
\lambda \rightarrow r^{i}(x,t)$ of (\ref{genek}). Then we obtain infinitely
many commuting $N$ component hydrodynamic type systems (here no summation
with respect to the index $i$)%
\begin{equation*}
r_{t_{k}}^{i}=a_{i}^{(k)}(\mathbf{r})r_{x}^{i},
\end{equation*}%
where (see (\ref{aka}))%
\begin{equation*}
a_{i}^{(k)}(\mathbf{r})=a^{(k)}(\lambda ,\mathbf{r})|_{\lambda =r^{i}}\equiv
(r^{i})^{k}+(r^{i})^{k-1}a_{1}+...+r^{i}a_{k-1}+a_{k}.
\end{equation*}%
However, for construction of $N$-phase solutions we can use just first $N-1$
commuting flows. These hydrodynamic type systems are linearly degenerate
(see, for instance, \cite{Makslin}, \cite{Fer91}), i.e. their characteristic
velocities $a_{i}^{(k)}(\mathbf{r})$ do not depend on corresponding Riemann
invariants ($\partial _{i}a_{i}^{(k)}(\mathbf{r})=0$, no summation here!).
Taking into account (\ref{separato}), we obtain the system%
\begin{equation}
r_{x}^{i}=2\frac{\sqrt{S(r^{i})}}{\prod_{m\neq i}(r^{i}-r^{m})},\text{ \ }%
r_{t_{k}}^{i}=2a_{i}^{(k)}(\mathbf{r})\frac{\sqrt{S(r^{i})}}{\prod_{m\neq
i}(r^{i}-r^{m})},\text{ }k=1,2,...,N-1.  \label{complet}
\end{equation}%
Taking into account (\ref{aka}) and (\ref{belpoli}), we obtain%
\begin{equation*}
a_{i}^{(k)}(\mathbf{r})=\tilde{a}_{i}^{(k)}(\mathbf{r})+\eta _{1}\tilde{a}%
_{i}^{(k-1)}(\mathbf{r})+...+\eta _{k-1}\tilde{a}_{i}^{(1)}(\mathbf{r})+\eta
_{k}.
\end{equation*}%
Introducing (instead of independent variables $t_{k}$) $N$ phases $\theta
_{k}$ such that (here $\eta _{0}=1$ and $t_{0}\equiv x$)%
\begin{equation}
\theta _{k}=\sum_{m=k}^{N-1}\eta _{m-k}t_{m},\text{ }k=0,1,...,N-1,
\label{teta}
\end{equation}%
system (\ref{complet}) becomes%
\begin{equation*}
r_{\theta _{k}}^{i}=2\tilde{a}_{i}^{(k)}(\mathbf{r})\frac{\sqrt{S(r^{i})}}{%
\prod_{m\neq i}(r^{i}-r^{m})},\text{ }k=0,1,2,...,N-1.
\end{equation*}%
This system also can be written in the Hamiltonian form\footnote{%
This\ Hamiltonian part is based on ideas of S. Alber, see, for instance, 
\cite{SAlber}. Later the same approach was applied in \cite{BM} for
multi-component dispersive reductions also considered in our paper.}%
\begin{equation*}
r_{\theta _{k}}^{i}=\frac{\partial H_{k}}{\partial \mu _{i}},\text{ \ }(\mu
_{i})_{\theta _{k}}=-\frac{\partial H_{k}}{\partial r^{i}},\text{ \ }%
i=1,2,...,N,\text{ \ }k=0,1,...,N-1,
\end{equation*}%
where Hamilton functions are (here $\tilde{a}_{i}^{(0)}(\mathbf{r})=1$)%
\begin{equation*}
H_{k}=\underset{i=1}{\overset{N}{\sum }}g_{(0)}^{ii}\tilde{a}_{i}^{(k)}(%
\mathbf{r})\mu _{i}^{2}+V_{k}(\mathbf{r})=\underset{i=1}{\overset{N}{\sum }}%
\frac{\tilde{a}_{i}^{(k)}(\mathbf{r})}{\prod_{m\neq i}(r^{i}-r^{m})}(\mu
_{i}^{2}-S(r^{i})).
\end{equation*}%
By virtue of the identities%
\begin{equation*}
\sum_{m=1}^{N}\frac{\partial r^{i}}{\partial \theta _{m}}\frac{\partial
\theta _{m}}{\partial r^{k}}=\delta _{k}^{i},\text{ \ }\sum_{m=1}^{N}\frac{%
\partial r^{m}}{\partial \theta _{i}}\frac{\partial \theta _{k}}{\partial
r^{m}}=\delta _{k}^{i},
\end{equation*}%
one can find all particular derivatives $\partial \theta _{i}/\partial r^{k}$%
. Their straightforward integration implies multi-phase solutions written in
the implicit form\footnote{%
This is nothing but the well-known Hamilton--Jacobi approach.}%
\begin{equation}
\theta _{k}=\frac{1}{2}\sum_{n=1}^{N}\int^{r^{n}}\frac{\lambda
^{N-k-1}d\lambda }{\sqrt{S(\lambda )}},\text{ }k=0,1,...,N-1.  \label{hj}
\end{equation}%
for integrable dispersive chains (\ref{mchains}) and their first $N-2$
commuting flows altogether, where (see (\ref{teta}))%
\begin{equation*}
t_{k}=\sum_{m=k}^{N-1}\kappa _{m-k}\theta _{m}.
\end{equation*}

\textbf{Remark}: Multi-phase solution (\ref{hj}) coincides with multi-phase
solution (\ref{gen}) for Mikhal\"{e}v system (\ref{three}) in the case (see (%
\ref{sryad}))%
\begin{equation}
g_{k}(\lambda )=2\sqrt{S(\lambda )},\text{ \ }k=0,1,...,N-1,  \label{link}
\end{equation}%
if we keep first three independent variables $x,t,y$ only (i.e. we fix
higher time variables to constants). The minor difference between both
constructions (the finite-gap integration and the method of hydrodynamic
reductions) appears in the one-phase and the two-phase solutions only.
Two-phase solution (\ref{hj}) for $M$th disperive chain (\ref{mchains}) is
(see (\ref{teta}))%
\begin{equation}
\theta _{0}=x+\eta _{1}t=\frac{1}{2}\sum_{n=1}^{2}\int^{r^{n}}\frac{\lambda
d\lambda }{\sqrt{S(\lambda )}},\text{ }\theta _{1}=t=\frac{1}{2}%
\sum_{n=1}^{2}\int^{r^{n}}\frac{d\lambda }{\sqrt{S(\lambda )}},  \label{a}
\end{equation}%
while the two-phase solution (see (\ref{gen})) for Mikhal\"{e}v system (\ref%
{three}) is (see (\ref{link}))%
\begin{equation}
\theta _{0}=x+\eta _{1}t+\eta _{2}y=\frac{1}{2}\sum_{n=1}^{2}\int^{r^{n}}%
\frac{\lambda d\lambda }{\sqrt{S(\lambda )}},\text{ \ }\theta _{1}=t+\eta
_{1}y=\frac{1}{2}\sum_{n=1}^{2}\int^{r^{n}}\frac{d\lambda }{\sqrt{S(\lambda )%
}}.  \label{b}
\end{equation}%
This means that two-phase solution (\ref{a}) of $M$th disperive chain (\ref%
{mchains}) can be obtained from two-phase solution (\ref{b}) of Mikhal\"{e}v
system (\ref{three}) by the stationary reduction $y\rightarrow $const.
One-phase solution of $M$th disperive chain (\ref{mchains}) can be obtained
directly from (\ref{gen})%
\begin{equation*}
\theta _{0}=x+\eta _{1}t+\eta _{2}y=\frac{1}{2}\int^{r^{1}}\frac{d\lambda }{%
g(\lambda )}
\end{equation*}%
by reduction (\ref{link}) and by the stationary reduction $y\rightarrow $%
const. In this particular case, plenty important formulas are simplified.
For instance, (\ref{raz}), (\ref{dva}) reduce to the form%
\begin{equation*}
u_{m}=\sum_{k=0}^{m}(k+1)s_{m-k}(r^{1})^{k},\text{ \ }1\leqslant m\leqslant
M;\text{ \ \ }u_{m}=\sum_{k=2}^{\infty }(k-1)s_{m+k}(r^{1})^{-k},\text{ \ }%
m>M.
\end{equation*}%
Also $a_{k}=\eta _{k}-\eta _{k-1}r^{1}$ and%
\begin{equation*}
\sigma _{s}=\sum_{m=0}^{s}\kappa _{m}(r^{1})^{s-m}.
\end{equation*}

\textbf{Remark}: Conservation law densities $\sigma _{k}(\mathbf{r})$ are
Bell polynomials. Indeed, (\ref{fifi}) can be rewritten in the form ($%
\varphi \rightarrow 1/p$, see (\ref{fi}) and (\ref{sigma}); $\Phi (\lambda ,%
\mathbf{r})\rightarrow 1/B(\lambda ,\mathbf{r})$, see (\ref{bryad}))%
\begin{equation*}
p(\lambda ,\mathbf{r})=\lambda ^{N}\left( 1+\frac{s_{1}}{\lambda }+\frac{%
s_{2}}{\lambda ^{2}}+\frac{s_{3}}{\lambda ^{3}}+...\right) ^{1/2}B(\lambda ,%
\mathbf{r}).
\end{equation*}%
Thus (we remind that $\kappa _{0}=1$ and $B_{0}=1$),%
\begin{equation*}
\sigma _{k}=\sum_{m=0}^{k}\kappa _{m}B_{k-m}.
\end{equation*}%
Taking into account relationships (\ref{svyaz}), all functions $a_{k}(%
\mathbf{r})$ can be also expressed via Bell polynomials, for instance,

\begin{equation*}
\begin{aligned} a_1 &= -B_1 - \tfrac{1}{2} s_1, \\ a_2 &= -B_2 + B_1^2 +
\tfrac{1}{2} s_1 B_1 - \tfrac{1}{2} s_2 + \tfrac{3}{8} s_1^2, \\ a_3 & =
-B_3 + 2 B_1 B_2 - B_1^3 + (\tfrac{1}{2} s_2 - \tfrac{3}{8} s_1^2) B_1 +
\tfrac{1}{2} s_1 (B_2 - B_1^2) \\&\quad - \tfrac{1}{2} s_3 + \tfrac{3}{4}
s_1 s_2 - \tfrac{5}{16} s_1^3, \\ a_4 &= -B_4 + 2 B_1 B_3 + B_2^2 - 3 B_1^2
B_2 + B_1^4 + (\tfrac{3}{8} s_1^2 - \tfrac{1}{2} s_2)(B_1^2 - B_2) \\&\quad
+ \tfrac{1}{2} s_1 (B_3 - 2 B_1 B_2 + B_1^3) + (\tfrac{1}{2} s_3 -
\tfrac{3}{4} s_1 s_2 + \tfrac{5}{16} s_1^3) B_1 \\&\quad - \tfrac{1}{2} s_4
+ \tfrac{3}{4} s_1 s_3 + \tfrac{3}{8} s_2^2 - \tfrac{15}{16} s_1^2 s_2 +
\tfrac{35}{128} s_1^4. \end{aligned}
\end{equation*}

\section{Multi-Component Dispersive Reductions}

\label{sec:multi}

In this Section we consider several integrable dispersive systems\footnote{%
Most of these integrable dispersive systems associated with the Energy
Dependent Schr\"{o}dinger operator were introduced in \cite{AF}, where their
multi-Hamiltonian structures were constructed. See also \cite{lma}.
Multi-phase solutions (\ref{hj}), (\ref{srac}) and their connection with the
Jacobi theta-function was investigated in \cite{Alber}.} which can be
extracted as dispersive reductions of dispersive chains (\ref{mchains}).

A wide class of integrable dispersive systems (see \cite{energy})%
\begin{equation}
u_{m,t}=u_{m+1,x}-u_{m}u_{1,x}-\frac{1}{2}u_{1}u_{m,x},\text{ \ }m=1,...,M-1,
\label{biggest}
\end{equation}%
\begin{equation*}
u_{M,t}=-u_{M}u_{1,x}-\frac{1}{2}u_{1}u_{M,x}+\overset{K}{\underset{k=1}{%
\sum }}w_{k,x}+\frac{1}{4}u_{1,xxx},
\end{equation*}%
\begin{equation*}
w_{k,t}=\left( \epsilon _{k}-\frac{1}{2}u_{1}\right) w_{k,x}-w_{k}u_{1,x},
\end{equation*}%
embedded into dispersive chains (\ref{mchains}) is selected by the rational
ansatz (see (\ref{you}))%
\begin{equation}
u=\lambda ^{M}+\overset{M}{\underset{m=1}{\sum }}\lambda ^{M-m}u_{m}+\overset%
{K}{\underset{k=1}{\sum }}\frac{w_{k}}{\lambda -\epsilon _{k}},
\label{racio}
\end{equation}%
where $\epsilon _{k}$ are pairwise distinct arbitrary constants and $w_{k}$
are new field variables, such that (see (\ref{you}))%
\begin{equation*}
u_{M+m}=\overset{K}{\underset{k=1}{\sum }}(\epsilon _{k})^{m-1}w_{k},\text{
\ }m=1,2,...
\end{equation*}

\textbf{Remark}: This rational ansatz (\ref{racio}) also can be written in
the factorised form%
\begin{equation*}
u=\frac{\prod_{m=1}^{M+K}(\lambda -q^{m})}{\prod_{k=1}^{K}(\lambda -\epsilon
_{k})},
\end{equation*}%
where $q^{m}$ are pairwise distinct field variables. Then integrable
dispersive systems (\ref{biggest}) take the form (see \cite{energy})%
\begin{equation*}
q_{t}^{i}=(q^{i}+a_{1})q_{x}^{i}+\frac{1}{2}\frac{\overset{K}{\underset{k=1}{%
\prod }}(q^{i}-\epsilon _{k})}{\underset{m\neq i}{\prod }(q^{i}-q^{m})}%
a_{1,xxx},
\end{equation*}%
where%
\begin{equation*}
a_{1}=\frac{1}{2}\left( \overset{M+K}{\underset{m=1}{\sum }}q^{m}-\overset{K}%
{\underset{k=1}{\sum }}\epsilon _{k}\right) .
\end{equation*}

In this generic case\footnote{%
The generic case means that the parameters $\epsilon _{k}$ are pairwise
distinct and the field variables $q^{k}$ are pairwise distinct.} Laurent
series (\ref{sryad}) reduces to the rational form%
\begin{equation}
S(\lambda )=\frac{\prod_{m=1}^{2N+M+K}(\lambda -\beta _{m})}{%
\prod_{k=1}^{K}(\lambda -\epsilon _{k})}.  \label{srac}
\end{equation}%
This means that corresponding multi-phase solutions of dispersive system (%
\ref{biggest}) are parameterised by a \textit{finite} number of arbitrary
constants $\beta _{k}$ only. The first $N$ functions $u_{m}(\mathbf{r})$ are
determined by (\ref{raz}), while the functions $w_{k}(\mathbf{r})$ take the
form%
\begin{equation*}
w_{k}=\frac{\prod_{m=1}^{2N+M+K}(\epsilon _{k}-\beta _{m})}{\underset{s\neq k%
}{\prod }(\epsilon _{k}-\epsilon _{s})}\prod_{n=1}^{N}(r^{n}-\epsilon
_{k})^{-2}
\end{equation*}%
\begin{equation*}
=\frac{\prod_{m=1}^{2N+M+K}(\epsilon _{k}-\beta _{m})}{\underset{s\neq k}{%
\prod }(\epsilon _{k}-\epsilon _{s})}\sum_{n=0}^{\infty }(\epsilon
_{k})^{n}\sum_{p=0}^{n}B_{-1-p}B_{-n-1+p}.
\end{equation*}

Plenty potentially interesting sub-cases can be investigated separately, for
instance, the such a case $\epsilon _{2}=\epsilon _{1}$. One can easily make
an appropriate computation. In this paper we omit detailed imvestigation of
infinitely many such particular cases. Here we just mention, that if all
parameters $\epsilon _{k}$ vanish, a simplest set of dispersive reductions
is selected by the constraint $u_{M+K+1}=0$, where $K=0,1,2,...$ In this
case (see (\ref{you}))%
\begin{equation*}
u=\lambda ^{M}\left( 1+\frac{u_{1}}{\lambda }+\frac{u_{2}}{\lambda ^{2}}+...+%
\frac{u_{M+K}}{\lambda ^{M+K}}\right) ,
\end{equation*}%
and Laurent series (\ref{sryad}) will be truncated and reduces to the
rational form (see (\ref{srac}))%
\begin{equation*}
S(\lambda )=\lambda ^{-K}\prod_{m=1}^{2N+M+K}(\lambda -\beta _{m}).
\end{equation*}

\textbf{Examples}:

1. If $K=0$, the equation of the Riemann surface is pure polynomial%
\begin{equation*}
\mu ^{2}=S(\lambda )=\prod_{m=1}^{2N+M}(\lambda -\beta _{m}).
\end{equation*}%
Thus, if $M=1$, the Korteweg--de Vries equation (\ref{kdv}) has $N$-phase
solutions (\ref{hj}), where (see (\ref{raz}))%
\begin{equation*}
u_{1}(\mathbf{r})=2\sum_{n=1}^{N}r^{n}-\sum_{m=1}^{2N+1}\beta _{m};
\end{equation*}%
if $M=2$, the Kaup--Boussinesq system (\ref{kdv}) has $N$-phase solutions (%
\ref{hj}), where (see (\ref{raz}))%
\begin{equation*}
u_{1}(\mathbf{r})=2\sum_{n=1}^{N}r^{n}-\sum_{m=1}^{2N+2}\beta _{m},
\end{equation*}%
\begin{equation*}
u_{2}(\mathbf{r})=\sum_{n=1}^{N}(r^{n})^{2}+2\left(
\sum_{n=1}^{N}r^{n}\right) ^{2}
\end{equation*}%
\begin{equation*}
-2\left( \sum_{m=1}^{2N+2}\beta _{m}\right) \sum_{n=1}^{N}r^{n}+\frac{1}{2}%
\left( \sum_{m=1}^{2N+2}\beta _{m}\right) ^{2}-\frac{1}{2}%
\sum_{m=1}^{2N+2}\beta _{m}^{2}.
\end{equation*}

2. If $K\geqslant 1$, the equation of the Riemann surface is rational%
\begin{equation}
\mu ^{2}=S(\lambda )=\lambda ^{-K}\prod_{m=1}^{2N+M+K}(\lambda -\beta _{m}).
\label{k}
\end{equation}%
Thus $N$-phase solutions (\ref{hj}) can be written in the form%
\begin{equation*}
\theta _{k}=\frac{1}{2}\sum_{n=1}^{N}\int^{r^{n}}\frac{\lambda
^{N+K/2-k-1}d\lambda }{\sqrt{\tilde{S}(\lambda )}},\text{ }k=0,1,...,N-1,
\end{equation*}%
where%
\begin{equation*}
\tilde{S}(\lambda )=\prod_{m=1}^{2N+M+K}(\lambda -\beta _{m}).
\end{equation*}%
For instance, if $M=1$ and $K=1$, the Ito system (\ref{ito}) has $N$-phase
solutions (\ref{hj})%
\begin{equation*}
\theta _{k}=\frac{1}{2}\sum_{n=1}^{N}\int^{r^{n}}\frac{\lambda
^{N-k-1/2}d\lambda }{\sqrt{\tilde{S}(\lambda )}},\text{ }k=0,1,...,N-1,
\end{equation*}%
where%
\begin{equation}
u_{1}(\mathbf{r})=2\sum_{n=1}^{N}r^{n}-\sum_{m=1}^{2N+2}\beta _{m},\text{ \ }%
u_{2}(\mathbf{r})=\left( \prod_{m=1}^{2N+2}\beta _{m}\right) \underset{n=1}{%
\overset{N}{\prod }}(r^{n})^{-2},  \label{itoh}
\end{equation}%
and%
\begin{equation*}
\tilde{S}(\lambda )=\prod_{m=1}^{2N+2}(\lambda -\beta _{m}).
\end{equation*}

\textbf{Remark}: In the generic case (\ref{srac}) constants $\kappa _{m}$
and $\eta _{m}$ can be expressed via $\beta _{k}$ as Bell polynomials, i.e. $%
\eta _{k}=C_{k}(\mathbf{\beta ,\epsilon })$, where%
\begin{equation}
C_{k}(\mathbf{\beta ,\epsilon })=\frac{1}{2k}\left(
\sum_{m=1}^{2N+M+K}(\beta _{m})^{k}-\sum_{m=1}^{K}(\epsilon _{m})^{k}\right)
.  \label{beta}
\end{equation}%
Indeed, taking into account (\ref{etas}), one can obtain (we remind that $%
\eta _{0}=1$ and $s_{0}=1$)%
\begin{equation*}
\sum_{m=0}^{\infty }\eta _{m}\lambda ^{-m}=\left( \sum_{m=0}^{\infty
}s_{m}\lambda ^{-m}\right) ^{-1/2}=\prod_{k=1}^{K}(1-\epsilon _{k}/\lambda
)^{1/2}\prod_{m=1}^{2N+M+K}(1-\beta _{m}/\lambda )^{-1/2}
\end{equation*}%
\begin{equation*}
=\exp \left[ \frac{1}{2}\sum_{k=1}^{K}\ln (1-\epsilon _{k}/\lambda )-\frac{1%
}{2}\sum_{m=1}^{2N+M+K}\ln (1-\beta _{m}/\lambda )\right]
\end{equation*}%
\begin{equation*}
=\exp \left[ \sum_{m=1}^{\infty }\frac{1}{2m}\left(
\sum_{k=1}^{2N+M+K}(\beta _{k})^{m}-\sum_{k=1}^{K}(\epsilon _{k})^{m}\right)
\lambda ^{-m}\right] .
\end{equation*}%
This means (see (\ref{beta}) and cf. (\ref{bel}))%
\begin{equation*}
\sum_{m=0}^{\infty }\eta _{m}\lambda ^{-m}=\exp \left( \sum_{m=1}^{\infty
}C_{m}\lambda ^{-m}\right) .
\end{equation*}%
Correspondingly (see (\ref{kapas})),%
\begin{equation*}
\sum_{m=0}^{\infty }\kappa _{m}\lambda ^{-m}=\exp \left( -\sum_{m=1}^{\infty
}C_{m}\lambda ^{-m}\right) .
\end{equation*}%
Thus, for all natural numbers $n$, we have 
\begin{equation*}
\begin{aligned} \eta_n &= \sum \frac{1}{k_1!\,k_2! \cdots k_m!}
C_{i_1}^{k_1} C_{i_2}^{k_2} \cdots C_{i_m}^{k_m}, \\ \kappa_n &= \sum
\frac{(-1)^{k_1 + k_2 + \cdots + k_m}}{k_1!\,k_2! \cdots k_m!} C_{i_1}^{k_1}
C_{i_2}^{k_2} \cdots C_{i_m}^{k_m}, \end{aligned}
\end{equation*}%
where the summation runs over all partitions%
\begin{equation*}
n=\underbrace{i_{1}+\cdots +i_{1}}_{k_{1}}+\underbrace{i_{2}+\cdots +i_{2}}%
_{k_{2}}+\cdots +\underbrace{i_{m}+\cdots +i_{m}}_{k_{m}}
\end{equation*}%
of the number $n$ as a sum of $m$ distinct positive natural numbers $%
i_{1},i_{2},\dots ,i_{m}$, where $m$ is arbitrary.

\section{The Exceptional Case $M=0$}

\label{sec:except}

In the particular case $M=0$ the compatibility conditions $(\psi
_{t_{1}})_{xx}=(\psi _{xx})_{t_{1}}$, $(\psi _{t_{2}})_{xx}=(\psi
_{xx})_{t_{2}}$ (see (\ref{sh}), (\ref{you}), (\ref{aka})) lead to
Camassa--Holm dispersive commuting chains%
\begin{equation}
u_{k,t_{1}}=u_{k+1,x}+a_{1}u_{k,x}+2u_{k}a_{1,x},  \label{zero}
\end{equation}%
\begin{equation*}
u_{k,t_{2}}=u_{k+2,x}+a_{1}u_{k+1,x}+2u_{k+1}a_{1,x}+a_{2}u_{k,x}+2u_{k}a_{2,x},
\end{equation*}%
where the functions $a_{1}$, $a_{2}$ are connected with $u_{1}$, $u_{2}$ via
($\xi _{1},\xi _{2}$ are arbitrary constants)%
\begin{equation*}
u_{1}=\frac{1}{2}a_{1,xx}-2a_{1}+\xi _{1},\text{ \ \ }u_{2}=\frac{1}{2}%
a_{2,xx}-2a_{2}-\frac{1}{2}a_{1}a_{1,xx}-\frac{1}{4}%
(a_{1,x})^{2}+3a_{1}^{2}-2\xi _{1}a_{1}+\xi _{2}.
\end{equation*}%
The simplest reduction $u_{2}=0$ yields the Camassa--Holm equation\footnote{%
Constants $\xi _{m}$ are not important. They can be removed by appropriate
Galilean transformation. For instance, in the case of the Camassa--Holm
equation, the constant $\xi _{1}$ can be removed by the transformation $%
z=x+\xi _{1}t/2$, $y=t$ and $a_{1}(x,t)=w(y,z)+\xi _{1}/2$.}%
\begin{equation}
u_{1,t_{1}}=a_{1}u_{1,x}+2u_{1}a_{1,x},\text{ \ }u_{1}=\frac{1}{2}%
a_{1,xx}-2a_{1}+\xi _{1}.  \label{ch}
\end{equation}%
In comparison with the general case $M>0$, in this exceptional case $M=0$,
the conservation law densities $\sigma _{k}$ no longer can be found from (%
\ref{fieks}) as differential polynomials with respect to the field variables 
$u_{m}$. Here we obtain $u_{m}$ as differential polynomials with respect to
the conservation law densities $\sigma _{k}$:%
\begin{equation*}
u_{1}=-\frac{1}{2}\sigma _{1,xx}+2\sigma _{1}+\xi _{1},\text{ \ }u_{2}=\frac{%
1}{2}(\sigma _{1}^{2}-\sigma _{2})_{xx}-\frac{1}{2}\sigma _{1}\sigma _{1,xx}-%
\frac{1}{4}(\sigma _{1,x})^{2}+\sigma _{1}^{2}+2\sigma _{2}+2\xi _{1}\sigma
_{1}+\xi _{2},
\end{equation*}%
\begin{equation*}
u_{3}=-\frac{1}{2}(\sigma _{3}-2\sigma _{1}\sigma _{2}+\sigma
_{1}^{3})_{xx}+2\sigma _{3}+2\sigma _{1}\sigma _{2}+\frac{3}{2}\sigma
_{1}(\sigma _{1,x})^{2}-\frac{1}{2}\sigma _{1,x}\sigma _{2,x}
\end{equation*}%
\begin{equation*}
+\left( \sigma _{1}^{2}-\frac{1}{2}\sigma _{2}\right) \sigma _{1,xx}-\frac{1%
}{2}\sigma _{1}\sigma _{2,xx}+\xi _{1}(\sigma _{1}^{2}+2\sigma _{2})-2\xi
_{2}\sigma _{1}+\xi _{3},...
\end{equation*}%
Moreover, in this exceptional case $M=0$, formula (\ref{raz}) is no longer
applicable. So, multi-phase solutions for the Camassa--Holm chains are
determined by (\ref{dva}) and (\ref{hj}), i.e.%
\begin{equation*}
u_{m}(\mathbf{r})=\sum_{k=1}^{\infty
}s_{2N+k+m-1}\sum_{n=1}^{k}B_{-n}B_{n-1-k},\text{ \ \ }\theta _{k}=\frac{1}{2%
}\sum_{n=1}^{N}\int^{r^{n}}\frac{\lambda ^{N-k-1}d\lambda }{\sqrt{S(\lambda )%
}},\text{ }k=0,1,...,N-1,
\end{equation*}%
where (see (\ref{sryad}))%
\begin{equation*}
S(\lambda )=\lambda ^{2N}\left( 1+\frac{s_{1}}{\lambda }+\frac{s_{2}}{%
\lambda ^{2}}+\frac{s_{3}}{\lambda ^{3}}+...\right) .
\end{equation*}%
Thus, multi-phase solutions for Camassa--Holm equation (\ref{ch}) take the
form (here $K=1$, cf. (\ref{itoh}))%
\begin{equation*}
u_{1}(\mathbf{r})=\left( \prod_{m=1}^{2N+1}\beta _{m}\right) \underset{n=1}{%
\overset{N}{\prod }}(r^{n})^{-2},\text{ \ \ }\theta _{k}=\frac{1}{2}%
\sum_{n=1}^{N}\int^{r^{n}}\frac{\lambda ^{N-k-1/2}d\lambda }{\sqrt{\tilde{S}%
(\lambda )}},\text{ }k=0,1,...,N-1,
\end{equation*}%
where (see (\ref{k}))%
\begin{equation*}
\tilde{S}(\lambda )=\prod_{m=1}^{2N+1}(\lambda -\beta _{m}).
\end{equation*}%
Camassa--Holm chain (\ref{zero}) also (as well as the general case $M>0$)
possesses two-dimensional reductions (cf. (\ref{biggest}))%
\begin{equation}
w_{k,t}=(\epsilon _{k}+a_{1})w_{k,x}+2w_{k}a_{1,x},  \label{w}
\end{equation}%
selected by rational ansatz (see (\ref{racio}))%
\begin{equation*}
u=1+\overset{K}{\underset{k=1}{\sum }}\frac{w_{k}}{\lambda -\epsilon _{k}},
\end{equation*}%
where the function $a_{1}$ is determined by the constraint%
\begin{equation*}
\overset{K}{\underset{k=1}{\sum }}w_{k,x}-\frac{1}{2}a_{1,xxx}+2a_{1,x}=0.
\end{equation*}%
This dispersive system (\ref{w}) possesses multi-phase solutions presented
in implicit form (cf. previous Section)%
\begin{equation*}
w_{k}=\frac{\prod_{m=1}^{2N+K}(\epsilon _{k}-\beta _{m})}{\underset{s\neq k}{%
\prod }(\epsilon _{k}-\epsilon _{s})}\prod_{n=1}^{N}(r^{n}-\epsilon
_{k})^{-2}=\frac{\prod_{m=1}^{2N+K}(\epsilon _{k}-\beta _{m})}{\underset{%
s\neq k}{\prod }(\epsilon _{k}-\epsilon _{s})}\sum_{n=0}^{\infty }(\epsilon
_{k})^{n}\sum_{p=0}^{n}B_{-1-p}B_{-n-1+p},
\end{equation*}%
\begin{equation*}
\theta _{k}=\frac{1}{2}\sum_{n=1}^{N}\int^{r^{n}}\frac{\lambda
^{N-k-1}d\lambda }{\sqrt{S(\lambda )}},\text{ }k=0,1,...,N-1,
\end{equation*}%
where (cf. (\ref{srac}))%
\begin{equation*}
S(\lambda )=\frac{\prod_{m=1}^{2N+K}(\lambda -\beta _{m})}{%
\prod_{k=1}^{K}(\lambda -\epsilon _{k})}.
\end{equation*}

\section{Conclusion}

\label{sec:final}

Three-dimensional linearly-degenerate Mikhal\"{e}v system (\ref{three})
possesses $N$ component two-dimensional hydrodynamic reductions
parameterised by $N$ arbitrary functions of a single variable. Each of them
possesses a general solution parameterised by $N$ arbitrary functions of a
single variable. However, just $N$ component two-dimensional linearly
degenerate hydrodynamic reductions possess global solutions parameterised by
an arbitrary number of constants. In this paper we describe connection of
global solutions of Mikhal\"{e}v system (\ref{three}) and multi-phase
solutions for integrable two-dimensional dispersive systems associated with
the energy dependent Schr\"{o}dinger operator.

Moreover, the finite-gap (or algebro-geometric) method of integration was
developed and applied for \textit{finite} component systems only. In this
paper we succesfully extended this approach to \textit{infinite} component
systems. For $M$th dispersive integrable chains associated with the Mikhal%
\"{e}v system we constructed infinitely many multi-phase solutions, which
depend on \textit{infinite} number of arbitrary parameters. Also we derived
the so called \textquotedblleft trace formulas\textquotedblright\ for
coefficients of the energy dependent Schr\"{o}dinger operator $u_{k}(\mathbf{%
r})$ as well as for conservation law densities $\sigma _{k}(\mathbf{r})$ and
for coefficients $a_{k}(\mathbf{r})$ of evolution with respect to higher
time variables. This approach allows to reconsider multi-phase solutions of
multi-component dispersive reductions from unified point of view.

\section*{Acknowledgements}

MM gratefully acknowledges the support from GA\v{C}R under project
P201/12/G028. MVP's work was partially supported by the grant of Presidium
of RAS \textquotedblleft Fundamental Problems of Nonlinear
Dynamics\textquotedblright\ and by the RFBR grant 17-01-00366. MVP also
thanks V.E. Adler, A.V. Aksenov, L.V. Bogdanov, Yu.V. Brezhnev, E.V.
Ferapontov, G.A. El, A.Ya. Maltsev, V.G. Marikhin, V.B. Matveev, A.E.
Mironov, A.O. Smirnov, A.I. Zenchuk for important discussions.

\end{document}